\documentclass[10pt,journal,comsoc]{IEEEtran}

\usepackage{cite}
\usepackage{amsmath,amssymb,amsfonts}
\usepackage[normalem]{ulem}
\usepackage{textcomp}
\usepackage{svg}      
\usepackage{xcolor}
\usepackage[most]{tcolorbox}
\usepackage[table]{xcolor}
\usepackage[caption=false,font=footnotesize]{subfig} 
\usepackage[hidelinks   ]{hyperref}
\usepackage{listings}
\usepackage{multirow}
\usepackage{graphicx}     
\usepackage{comment}
\usepackage{pifont}
\usepackage{booktabs}
\usepackage{colortbl}
\usepackage{array}

\usepackage[nolist]{acronym}
\acrodef{CTM}{Convolutional Tsetlin Machine}
\acrodef{CNN}{Convolutional Neural Network}
\acrodef{TM}{Tsetlin Machine}
\acrodef{TA}[TA]{Tsetlin automaton}
\acrodefplural{TA}[TAs]{Tsetlin automata}
\acrodef{LA}[LA]{learning automaton}
\acrodefplural{LA}[LAs]{learning automata}
\acrodef{FPGA}{field-programmable gate array}
\acrodef{LUT}{look-up table}
\acrodef{BRAM}{block RAM}
\acrodef{SSB}{Synchronization Signal Block}
\acrodef{PSS}{Primary Synchronization Signal}
\acrodef{SSS}{Secondary Synchronization Signal}
\acrodef{PBCH}{Physical Broadcast Channel}
\acrodef{NR}{New Radio}
\acrodef{DNN}{Deep Neural Network}
\acrodef{NN}{Neural Network}
\acrodef{ML}{machine learning}
\acrodef{FFT}{Fast Fourier Transform}
\acrodef{OFDM}{orthogonal frequency-division multiplexing}
\acrodef{PCI}{Physical Cell Identity}
\acrodef{EVM}{error vector magnitude}
\acrodef{RSSI}{Received Signal Strength Indicator}
\acrodef{SJNR}{signal-to-jamming-plus-noise ratio}
\acrodef{AUC}{area under the ROC curve}
\acrodef{CFO}{carrier frequency offset}
\acrodef{TO}{time offset}
\acrodef{relu}[ReLU]{rectified linear unit}
\acrodef{ue}[UE]{user equipment}
\acrodef{gnb}[gNB]{next-generation NodeB}
\acrodef{DT-DDNN}{Deep Transformer-Driven Deep Denoising Neural Network}
\acrodef{Z7}{Zybo Z7: Zynq-7000}

\tcbset{
  colback=white, colframe=black!35,
  boxsep=1pt, left=4pt, right=4pt, top=3pt, bottom=3pt,
  before skip=4pt, after skip=4pt,
  fonttitle=\bfseries\footnotesize
}

\usepackage{algorithm}
\usepackage{algpseudocode}

\makeatletter
\newcommand{\algmargin}{\the\ALG@thistlm}
\makeatother

\newlength{\forwidth}
\algdef{SE}[parFOR]{parFor}{EndparFor}[1]
{\parbox[t]{\dimexpr\linewidth-\algmargin}{%
\hangindent\forwidth\strut\algorithmicfor\ #1\ \algorithmicdo\strut}}{\algorithmicend\ \algorithmicfor}%

\newlength{\whilewidth}
\algdef{SE}[parWHILE]{parWhile}{EndparWhile}[1]
{\parbox[t]{\dimexpr\linewidth-\algmargin}{%
\hangindent\whilewidth\strut\algorithmicwhile\ #1\ \algorithmicdo\strut}}{\algorithmicend\ \algorithmicwhile}%
\algnewcommand{\parState}[1]{ %
\parbox[t]{\dimexpr\linewidth-\algmargin}{\strut #1\strut}}

\usepackage{epstopdf}

\begin{document}

\title{Explainable and Hardware-Efficient Jamming Detection for 5G Networks Using the Convolutional Tsetlin Machine}
\author{
\IEEEauthorblockN{Vojtech Halenka, Mohammadreza Amini~\IEEEmembership{SMIEEE}, Per-Arne Andersen, Ole-Christoffer Granmo, and Burak Kantarci~\IEEEmembership{SMIEEE}\vspace{-0.1in}}\\
\thanks{V. Halenka, P-A. Andersen and O-C. Granmo, are with the Center of Artificial Intelligence Research, University of Agder, Grimstad, Norway. Emails: \{vojtech.halenka,per.andersen\}@uia.no\\
M. Amini and B. Kantarci are with the School of Electrical Engineering and Computer Science, University of Ottawa, Ottawa, ON, Canada. Emails: \{mamini6,burak.kantarci\}@uottawa.ca\\}}


\IEEEtitleabstractindextext{
\begin{abstract}
All applications in fifth-generation (5G) networks rely on stable radio-frequency (RF) environments to support mission-critical services in mobility, automation, and connected intelligence. Their exposure to intentional interference or low-power jamming threatens availability and reliability, especially when such attacks remain below link-layer observability. This paper investigates lightweight, explainable, and hardware-efficient jamming detection using the Convolutional Tsetlin Machine (CTM) operating directly on 5G Synchronization Signal Block (SSB) features. CTM formulates Boolean logic clauses over quantized inputs, enabling bit-level inference and deterministic deployment on FPGA fabrics. These properties make CTM well suited for real-time, resource-constrained edge environments anticipated in 5G. The proposed approach is experimentally validated on a real 5G testbed using over-the-air SSB data, emulating practical downlink conditions. We benchmark CTM against a convolutional neural network (CNN) baseline under identical preprocessing and training pipelines. On the real dataset, CTM achieves comparable detection performance (Accuracy 91.53 ± 1.01 vs.\ 96.83 ± 1.19 for CNN) while training $9.5\times$ faster and requiring $14\times$ less memory (45~MB vs.\ 624~MB). Furthermore, we outline a compact FPGA-oriented design for Zybo~Z7 (Zynq-7000) and provide resource projections (not measured) under three deployment profiles optimized for latency, power, and accuracy trade-offs. The results show that the CTM provides a practical, interpretable, and resource-efficient alternative to conventional DNNs for RF-domain jamming detection, establishing it as a strong candidate for edge-deployed, low-latency, and security-critical 5G applications while laying the groundwork for B5G systems.

\end{abstract}

}
\pagestyle{empty}

\maketitle
\thispagestyle{empty}
\IEEEdisplaynontitleabstractindextext
\IEEEpeerreviewmaketitle

\section{Introduction}


The increasing reliance of mission-critical services on the wireless medium inevitably exposes them to intentional radio-frequency (RF) interference (i.e., \emph{jamming}), which can severely degrade network availability and service reliability. Traditional link-layer monitoring and performance counters often fail to capture such attacks, particularly when jammers operate at low power levels or exploit dynamic and fast-fading channel conditions to remain undetected~\cite{khan2019survey,pirayesh2022survey,lichtman2018nrjamming}. 

Despite the growing volume of jamming detection work, many state-of-the-art solutions depend on higher-layer KPIs or deep models whose memory footprint, training cost, and limited transparency hinder deployment on IoT edge platforms. At the same time, the 5G SSB and PSS offer a deterministic, frequently transmitted physical-layer anchor whose distortions can reveal interference even when link-layer counters remain inconclusive. Motivated by this, we formulate RF-domain jamming detection direcly from over-the-air SSB features and adopt the CTM as a lightweight, logic-based alternative to conventional CNNs. By learning Boolean clauses over quantized time-frequency patterns, the proposed detector couples interpretability with an on-edge deployable model, enabling a practical accuracy-efficiency trade-off for security-critical 5G networks.

 The contributions of this paper are as follows.
\begin{enumerate}
  \item Develop a robust \ac{CTM}-based jamming detection framework that operates directly on over-the-air 5G signal features, eliminating the need for higher-layer performance indicator checks. The proposed detector is fully explainable and hardware-efficient, making it well-suited for deployment on resource-constrained IoT platforms.
  \item Benchmark \ac{CTM} against a strong \ac{CNN} baseline on the same features. While the \ac{CNN} attains slightly higher accuracy, \ac{CTM} trains an order of magnitude faster with a $\sim$14$\times$ smaller memory footprint, making it appealing for frequent updates and embedded deployment.
 \item Provide projections for \ac{Z7} board based on prior \ac{CTM}-on-\ac{FPGA} reports, outlining three deployment profiles with a different focus (Power, Latency, Accuracy).
\end{enumerate}

Our results complement and extend recent \ac{SSB}-centered jamming studies~\cite{flores2023uplinkjam,asemian2024mobility, asemian2025dtddnn, amini2025deepfusion}, by showing that \ac{ML} can approach \ac{CNN} accuracy on the same RF-domain features while significantly reducing training cost and model size—an attractive trade-off where power, memory, and deterministic latency matter.

The remainder of this paper is organized as follows. 
Section \ref{Sec:system model} describes the system model and problem formulation, including the 5G~\ac{NR} \ac{SSB} structure and jamming hypotheses. 
Section \ref{Sec: CTM-based} details the proposed \ac{CTM}-based jamming detection framework, covering preprocessing, Booleanization, and model configuration, along with the \ac{CNN} baseline for comparison. 
Section \ref{Sec:Result} presents the experimental results and performance analysis. Finally, Section \ref{Sec:Conclusion} concludes the paper with a summary of findings and future research directions.

\section{Related work}

A substantial body of literature has investigated 5G jamming detection mechanisms across different protocol layers. At the physical layer, researchers have exploited variations in signal-level statistics to infer abnormal interference patterns. These include the \ac{EVM}, \ac{RSSI}, and time–frequency spectrograms~\cite{arjoune2020hoeffding,ornek2022menacomm,chiarello2021pimrc,feng2018mlrfjam}. Complementary studies have analyzed the impact of jamming on the cell-search and synchronization process itself, using behavioral deviations as detection cues~\cite{chen2020initialsearch,tuninato2023syncstudy}. In parallel, a growing trend in the literature focuses on data-driven detection, leveraging supervised and unsupervised learning methods such as \ac{DNN}s, autoencoders, and attention-based models~\cite{jere2023twc,wang2022hpsr}. These approaches often achieve excellent classification accuracy and robustness against complex interference conditions. However, their reliance on large-parameter models and high-cost training/inference pipelines significantly limits their feasibility for deployment in real-time, resource-constrained edge or vehicular platforms. 

In the 5G \ac{NR} air interface, the \ac{SSB} represents the essential downlink structure that supports the earliest stages of cell search and initial access~\cite{3gpp38211,lin2018sspbcgws}. Because of its deterministic form, well-defined periodicity, and fixed mapping to the time–frequency grid, the \ac{SSB} not only plays a pivotal role in establishing connectivity but also becomes an attractive target. Even selective or low-power jammers can impair the correlation of the \ac{PSS} and \ac{SSS} or corrupt the decoding of the \ac{PBCH}. Such impairments can obstruct \ac{PCI} discovery or system information decoding while keeping the jammer's activity stealthy and energy-efficient~\cite{ludant2021sigunder,lichtman2018nrjamming,wang2023pbch_spca}. 

To overcome these limitations, several studies have introduced lightweight machine-learning frameworks specifically tailored for 5G jamming detection. For example, the works in~\cite{amini2025deepfusion,asemian2025dtddnn} proposed advanced detectors based on a Trust Evaluation Data Analysis (TEDA) module and a \ac{NN} \ac{DT-DDNN}, respectively. Both methods demonstrated strong detection capability and low false-alarm rates even under challenging \ac{SJNR} regimes. Nonetheless, despite their algorithmic elegance, the hardware and computational demands of TEDA and \ac{DT-DDNN} remain prohibitive for practical integration into compact or embedded platforms. 

\begin{figure}[!htbp]
\centering
\includegraphics[width=.9\linewidth]{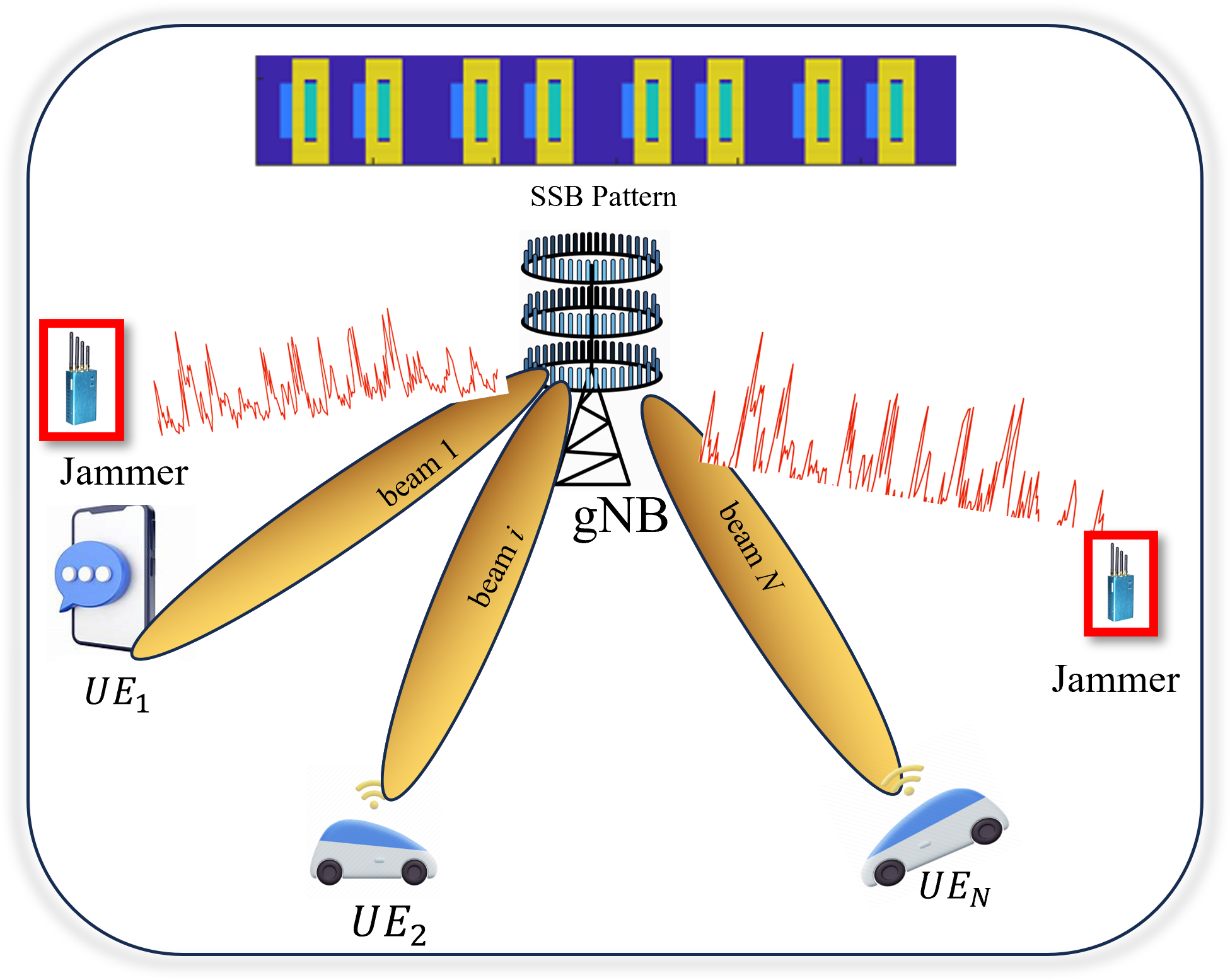}\vspace{-0.2in}
\caption{Network Model in consideration\vspace{-0.2in}}
\label{fig:system model}
\end{figure}

Motivated by these operational constraints, this work revisits \ac{SSB}-domain jamming detection through the lens of \emph{\ac{TM}s} and their convolutional extension, the \emph{\ac{CTM}}~\cite{granmo2018tm,granmo2019convolutional,abeyrathna2019weightedtm}. Inspired by the successful application in similar domain~\cite{Jeeru2025, Jeeru2025B}, we adapt the \ac{TM} to this use case. \ac{TM}s constitute a symbolic learning paradigm that formulates propositional logic clauses over Booleanized feature vectors using ensembles of \acp{LA}. This results in models that are simultaneously interpretable, statistically efficient, and highly suitable for low-bit or integer-based implementation. Unlike conventional floating-point \ac{DNN}s, \ac{TM} and \ac{CTM} inference operates via deterministic bit-wise logic, which naturally aligns with \ac{LUT} and \ac{BRAM} dominated \ac{FPGA} architectures, as shown in~\cite{Tunheim2025, Tunheim2025b}. Consequently, they enable fully on-chip models characterized by low power consumption, predictable latency, and hardware-deterministic behavior. Moreover, recent \ac{FPGA}-centric design flows—such as clause specialization, hierarchical clause sharing, and optimized bitstream synthesis—have further improved the area efficiency and scalability of large-clause \ac{TM} deployments~\cite{rule2024matador}.

\section{System model and problem formulation} \label{Sec:system model}

Consider a 5G network where a \ac{gnb} is located at the cell center, providing downlink services to multiple \acp{ue} distributed across the coverage area. In this environment, several malicious jammers attempt to disrupt the RF-domain transmissions, as depicted in Fig.~\ref{fig:system model}. According to the \ac{SSB} pattern specified in the 3GPP standard \cite{3gpp38211}, the \ac{gnb} periodically broadcasts synchronization signals across the network to enable \acp{ue} to perform cell search, timing acquisition, and initial access.
In 5G \ac{NR}, the \ac{PSS} serves as the key element for initial cell search, sector identification, and time–frequency synchronization at the \ac{ue} side. It constitutes the first \ac{OFDM} symbol of the \ac{SSB} and is constructed from one of three predefined $m$-sequences $\Phi_{N_{ID}^{(2)}}(k)$, each corresponding to a sector identity $N_{ID}^{(2)} \in \{0,1,2\}$. These sequences are generated according to a linear feedback shift register (LFSR) recurrence relation that defines a length-$k_{pss}=127$ pseudo-random binary sequence $s(i)$ as \vspace{-3mm}

\begin{equation}\label{eq: sequence} 
\begin{split}
  s(i+7)=(s(i+4)+s(i)&) \bmod 2\\
 [s(6) \ s(5) \ s(4) \ s(3) \ s(2) \ s(1) \ s(0)] &= [1 \ 1 \ 1 \ 0 \ 1 \ 1 \ 0]
\end{split}
\end{equation} 

\noindent where $s(i)\in\{0,1\}$.
The complex-valued PSS sequence in the frequency domain is then generated from the binary sequence as,
\begin{equation} \label{eq: m-sequences}
 \begin{split}
 \Phi_{N_{ID}^{(2)},k}&=1-2s(m)\\
 m=(k+43\times N^{(2)}_{ID})&\bmod k_{pss},0\leq k<k_{pss}
 \end{split}
\end{equation}

\noindent ensuring that each $\Phi_{N_{ID}^{(2)}}(k)$ is a circularly shifted version of the other two base sequences.

In the 5G resource grid, the \ac{PSS} occupies the central $127$ subcarriers of the first \ac{OFDM} symbol ($l=0$) within the \ac{SSB} bandwidth, while the subcarriers outside this range are nulled to maintain spectral separation \cite{3gpp38211}. The \ac{PSS} in frequency domain is written as
\begin{equation}
 X_{l,k}^{ssb} =
 \begin{cases}
 \Phi_{N_{ID}^{(2)},k}, & k \in \{56,57,\ldots,182\},\\
 0, & \text{otherwise}.
 \end{cases}
\end{equation}

\noindent After mapping the \ac{PSS} to its respective subcarriers, the time-domain representation of the \ac{OFDM} symbol is obtained through the inverse \ac{FFT}: \vspace{-3mm}

\begin{equation}
 x^{ssb}_l(n) = 
 \frac{1}{k_{ssb}} 
 \sum_{k=0}^{k_{ssb}-1} 
 X_{l,k}^{ssb}
 e^{j\frac{2\pi}{k_{ssb}}nk}.
\end{equation}

The transmitted signal $x^{ssb}_l(n)$ then propagates through the wireless channel, where it experiences the channel impulse response $h(n)$ and additive thermal noise $z(n)$. Consequently, the received \ac{SSB} at the user terminal can be modeled as
\begin{equation}
 y^{ssb}_l(n) 
 = x^{ssb}_l(n)\ast h(n) + z(n).
\end{equation} 
where $N_{\text{FFT}}$ is the \ac{FFT} length.

Since the \ac{PSS} is transmitted in the first \ac{OFDM} symbol ($l=0$), the received \ac{PSS} signal can be extracted directly from the first $N_{\text{FFT}}$ samples of the received \ac{SSB} as
\begin{equation}
 y_{pss}(n) = y^{ssb}_l(n)\big|_{l=0}, 
 \quad n \in \{0,\ldots,N_{\text{FFT}}-1\}.
\end{equation}

The extracted \ac{PSS} sequence $y_{pss}(n)$ serves as the basis for synchronization, channel estimation, and, in this work, jamming detection. The deterministic structure of the \ac{PSS}, its fixed location in the resource grid, and its excellent autocorrelation properties make it an ideal reference for analyzing interference-induced distortions in the RF domain.

Under the null hypothesis \( H_0 \), the received signal without jamming is, \vspace{-3mm}

\begin{equation}
 H_0 : y_{pss}(n) = x(n) \ast h(n) + z(n)
\end{equation}
Under the alternative hypothesis \( H_1 \), where the jamming signal ($x_j(n)$) is present, the received \ac{PSS} is
\begin{equation}
 H_1 : y_{pss}(n) = x(n) \ast h(n) + z(n) + x_j(n)
\end{equation}
The objective is to design a detection algorithm that discriminates between the two hypotheses and thus determines the presence or absence of a jamming attack. The receiver first extracts the \ac{PSS} time–frequency representation from the recorded IQ samples by applying a short-time \ac{FFT} and then constructs the input tensor for the \ac{CTM}-based detector. This pipeline yields an explainable, hardware-efficient detector that remains effective against both constant and sophisticated jammers, including under high \ac{SJNR} conditions.

\section{Proposed CTM-based jammer detection}\label{Sec: CTM-based}

In this section, we present the proposed approach for jammer detection in 5G networks using the \ac{CTM}. 

\subsection{Data Capturing and Preprocessing}

The acquired over-the-air IQ samples undergo several pre-processing stages prior to feature extraction. 
First, raw base-band signals are captured by the receiver.
To enable proper synchronization with the serving \ac{gnb}, both \ac{CFO} and \ac{TO} are estimated and compensated. 
Because the exact center frequency of the \ac{gnb} is unknown, a blind search procedure is applied. 
The \ac{CFO} is estimated by maximizing the correlation between the received and locally generated \ac{SSB} waveforms as

\begin{equation}
 \hat{f}_{\mathrm{CFO}}
 = \arg\max_{f_i}
 \Bigg[
 \sum_{\tau}
 y(\tau)
 e^{j2\pi \frac{f_i}{f_s}\tau}
 x^{ssb}_l(t-\tau)\Big|_{l=0}
 \Bigg].
\end{equation}
Next, the Schmidl–Cox method \cite{Schmidl1997} is adopted to determine the symbol-level timing offset by exploiting the cyclic prefix (CP) of the CP-\ac{OFDM} waveform as, \vspace{-5mm}

\begin{equation}\label{eq: optimization TO}
\begin{split}
 \hat{t}_{off} &= \arg\max_{t} \, M(t)=\frac{|P(t)|^2}{R(t)^2} ,
\end{split}
\end{equation}
in which $P(t)$ and $R(t)$ are presented as below: \vspace{-5mm}

\begin{equation}\label{eq: TO-P}
\begin{split}
 P(t)=\sum_{m=0}^{L-1} y^*(t+m)y(t+m+L)
\end{split}
\end{equation}

\begin{equation}\label{eq: TO-R}
\begin{split}
 R(t)=\sum_{m=0}^{L-1}|y(t+m+L)|^2 
\end{split}
\end{equation}
where $L$ is CP length.
After synchronization, the compensated waveform is aligned to the start of the \ac{SSB}. 
The first \ac{OFDM} symbol ($l=0$) is extracted to isolate \ac{PSS} for subsequent feature computation. This preprocessing chain ensures that the received signal is accurately frequency- and time-aligned with the \ac{gnb} transmission, providing a clean input for the proposed for both the \ac{CNN} and \ac{CTM}-based jamming detector.

Next, we preprocess the data for \ac{TM}-based classification. We tested several binarization techniques, resulting in sets of booleans, where each value represents a specific feature of the signal at a specific level, such as the presence or absence of a particular frequency or time pattern. This step is necessary for the \ac{TM}, which operates on Boolean features. 

\subsection{CTM for Jamming Detection}

Once the data is prepared and preprocessed, we apply the \ac{CTM} \cite{granmo2019convolutional} for the classification task, and compare with a conventional \ac{CNN}.

\begin{figure}[!htbp]
\centering \vspace{-3mm}
\includegraphics[width=1\linewidth]{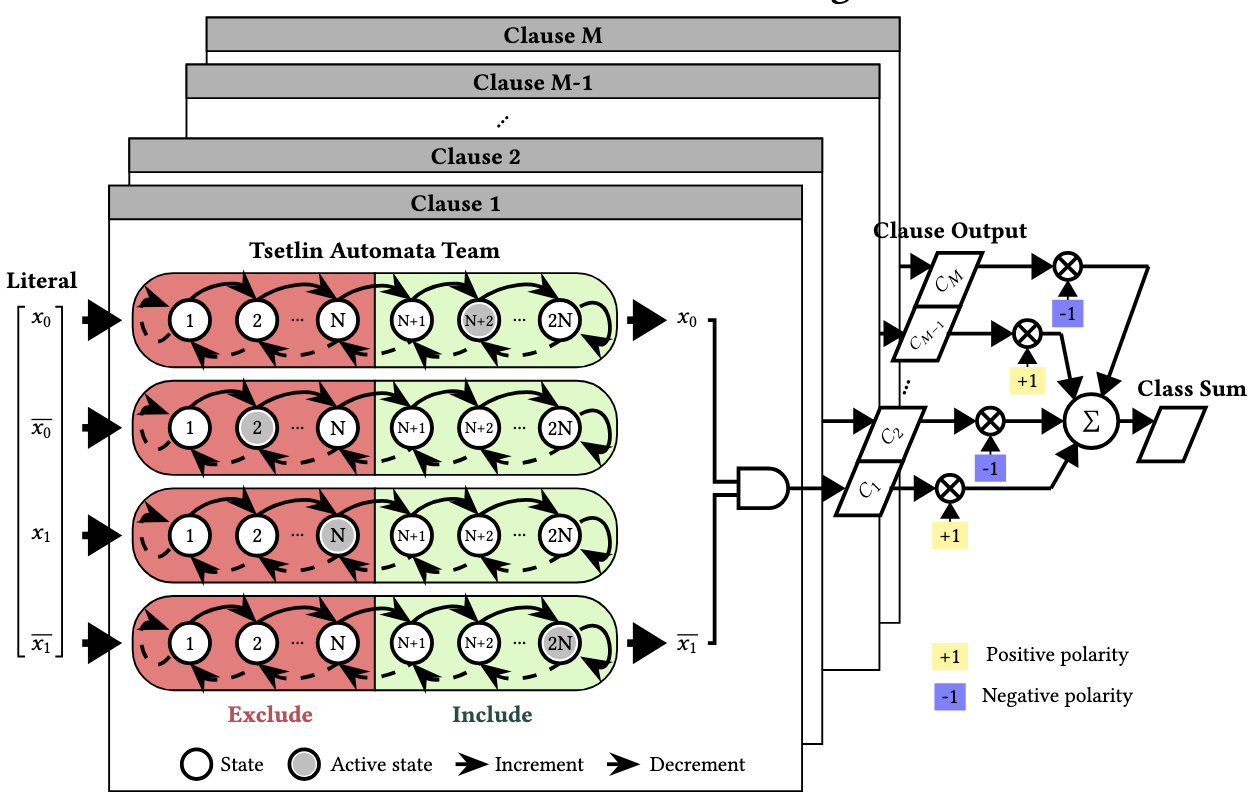}
\caption{The \ac{TM} structure \cite{duan2025}\vspace{-0.2in}}
\label{fig:TM} \vspace{-3mm}
\end{figure}

In short, \ac{TM} is a machine learning model based on propositional logic that uses a set of \acp{TA} to learn patterns in boolean data. As seen in Fig~\ref{fig:TM}, the key advantages of \ac{TM} arise from its native simplicity, and manifest as efficiency, transparency, and speed of classification of the model. For a thorough description of \ac{TM}, see~\cite{granmo2018tm} and for \ac{CTM}, see~\cite{granmo2019convolutional}. The \ac{CTM} expands \ac{TM} by processing the input data in patches of predefined size, showing a convolutional view to the model. The processing includes attaching coordinates to each patch.

\subsection{CTM Parameters}
We use \ac{CTM} with the configuration listed below. Values were tuned with Optuna~\cite{optuna2019} using 50 trials per preprocessing method on a validation subset and kept for all main results. The hyperparameter search space was
\vspace{-0.2in}

\begin{align*}
\texttt{number\_of\_clauses} &\in \{200\} \\
\texttt{T} &\in [50, 1000] \\
\texttt{s} &\in [1.0, 20.0] \\
\texttt{patch\_dim} &\in \{5{\times}5, .. 10{\times}10\} \\
\texttt{max\_included\_literals} &\in [20, 64] \\
\texttt{epochs} &\in [3, 10]
\end{align*}

The best performing configuration on the validation set was with Enhanced Otsu method and the following parameters: \vspace{-0.1in}

\begin{align*}
\texttt{TMClassifier(} 
&\texttt{number\_of\_clauses = 200,} \\
&\texttt{T = 477,} \texttt{s = 2.081,} \\
&\texttt{patch\_dim = (10, 10),} \\
&\texttt{max\_included\_literals = 22)}
\end{align*}
 
\subsection{CNN Baseline}
For direct comparison with \ac{CTM}, we trained a \ac{CNN}, inspired by~\cite{asemian2025dtddnn}, on the same preprocessed spectrograms, reshaped to $(H,W,1)$ with $H=W=100$. The architecture comprises four convolutional blocks with progressively decreasing filter counts (256, 256, 128, 64), followed by two dense layers and a softmax head. Each convolutional block uses a $3\times3$ convolution with \ac{relu} activation, batch normalization, $2\times2$ max-pooling, and dropout (rate 0.25). The fully-connected stack consists of two dense layers with 512 and 256 neurons respectively, each followed by batch normalization and dropout (rate 0.5). The final layer is a 2-way softmax. We trained with Adam (learning rate $10^{-3}$) and categorical cross-entropy loss, employing early stopping on validation accuracy with patience of 15 epochs.

\paragraph{Model Complexity and Experimental Fairness}
The \ac{CNN} baseline contains approximately 1.2~M trainable parameters and requires 45.3~M FLOPs per inference. To ensure fair comparison, both \ac{CTM} and \ac{CNN} were trained on identical splits of the dataset and preprocessed inputs. The \ac{CTM} operates on binarized spectrograms (Enhanced Otsu thresholding), while the \ac{CNN} uses normalized floating-point inputs. As an ablation study, we also evaluated the \ac{CNN} on binarized inputs, which resulted in significant performance degradation (accuracy: 93.71 → 74.69), confirming that \ac{CNN}s benefit from continuous-valued features.

\section{Experimental results} \label{Sec:Result}
 
The raw baseband data within the 5G FR1 n71 band are captured using the ThinkRF RTSA R5500 spectrum analyzer. Controlled jamming is realized by injecting a synthesized interference waveform through a wireline RF combiner at the receiver front end, where it is superimposed on an over-the-air captured 5G signal. This experimental design allows precise control over jammer power and waveform characteristics while ensuring regulatory compliance by preventing over-the-air jamming transmissions. The acquired signals are recorded in CSV format via the PyRF4 API. Sampling is performed at a rate of $15.625$~MHz with a carrier frequency of $f_c = 632$~MHz, corresponding to the Telus downlink center frequency. This configuration provides adequate spectral and temporal resolution for accurate synchronization and subsequent signal analysis.\footnote{Since the n71 band is shared between two operators, the selected sampling rate satisfies the bandwidth requirements specified in \cite{3gpp38211}.} The time--frequency features of the \ac{PSS}, obtained via \ac{FFT}, are then processed through the \ac{CTM} pipeline. The complete set of experimental parameters is summarized in Table~\ref{tab:exp-setup}. It is noted that the jammer signal, with multiple power levels as listed in Table~\ref{tab:exp-setup}, is combined with the received 5G signal using an RF combiner.

\begin{table}[!htbp]
\centering
\caption{Setup and acquisition parameters for data collection.}
\label{tab:exp-setup}
\small
\begin{tabular}{lc}
\toprule
\textbf{Parameter} & \textbf{Value} \\
\midrule
Carrier Frequency ($f_c$) & 632~MHz \\
Bandwidth & 35~MHz \\
Sampling Rate & 15.625~MHz \\
Antenna Type & Wideband omnidirectional \\
Duration per Capture & 200~ms \\
FFT Size & 1024 \\
Jammer Transmit Gain & (-80 to -40)~dB, step: 2~dB \\
Jammer Output Power & 5~dBm \\
Receiver Gain & 40~dB \\
CP Length & Normal  \\
\bottomrule
\end{tabular}
\end{table}

We evaluate \ac{CTM} with various preprocessing methods against a \ac{CNN}, on the same dataset with 5-fold cross-validation split. Both ran on the same hardware, Mac M1 Pro 16GB 2021. Table~\ref{tab:tm-cnn} summarizes accuracy-oriented and model-oriented metrics. The \ac{CNN} attains higher accuracies, while \ac{CTM} offers markedly lower training cost and smaller memory footprint, making it attractive for rapid iteration and resource-constrained deployments (e.g., \ac{FPGA} prototyping).

\paragraph{Booleanization}

We compared four binarization methods for \ac{CTM} preprocessing: Enhanced Otsu (Otsu thresholding applied to both the original image and its 90°-rotated version, combined via logical OR), Paper Original (global standard-deviation-based denoising followed by mean-value thresholding), Multi-Level (Otsu thresholding at full 100×100 resolution and half 50×50 resolution, upsampled and combined via OR), and Adaptive Threshold (local Gaussian adaptive thresholding with block size 11×11). Enhanced Otsu achieved the best accuracy (91.75 ± 0.83) by capturing both horizontal and vertical edge patterns critical for jamming signal detection. See results of achieved accuracy scores using \ac{CTM} in a 5-fold cross-validation study in Table \ref{tab:CTM_stats}.

\begin{table}[!htbp]
\centering
\caption{Preprocessing methods: 5-fold cross-validation accuracy with \ac{CTM} best found parameters}
\label{tab:CTM_stats}
\begin{tabular}{llllll}
\toprule
\textbf{Preprocessing} & \textbf{Accuracy} & \textbf{s} & \textbf{T} & \textbf{patch} & \textbf{max\_lit}  \\
\midrule
Enhanced Otsu        & \textbf{91.53 ± 1.01}       & 2.08     & 477       & 10×10      & 22            \\
Paper Original       & 91.47 ± 1.06       & 3.64     & 326       & 10×10      & 28            \\
Multi Level          & 90.88 ± 1.09       & 1.06     & 721       & 7×7        & 50            \\
Adaptive Thresh   & 80.20 ± 1.64       & 1.78     & 192       & 9×9        & 30            \\
\ac{CNN}             & \textbf{96.83 ± 1.19}       & --       & --        & --         & --             \\
\bottomrule 
\end{tabular}
\end{table}

The \ac{CNN} has shown only loss in performance when used with the same preprocessing techniques.

\paragraph{Accuracy}
The \ac{CNN} achieves higher scores than \ac{TM} in almost all metrics, indicating stronger discriminative performance, as shown in Table~\ref{tab:CTM_VS_CNN_stats}. However, there is a cost to that, in terms of memory and time.

\begin{table}[!htbp]
\centering
\caption{Accuracy and complexity metrics of CTM vs CNN}
\label{tab:CTM_VS_CNN_stats}
\setlength{\tabcolsep}{4pt} 
\small
\begin{tabular}{lcccc}
\toprule
\textbf{Model} 
& \textbf{Acc. [\%]} 
& \textbf{Train. Time [s]} 
& \textbf{Infer. Time [s]} 
& \textbf{Mem. [MB]} \\
\midrule
CTM 
& 91.96 
& \textbf{44.4} 
& 2.5 
& \textbf{45.2} \\
CNN 
& \textbf{96.81} 
& 3205.4 
& \textbf{0.8} 
& 892.1 \\
\bottomrule
\end{tabular} \vspace{-3mm}
\end{table}


\paragraph{Training cost}
\ac{TM} trains about $9.6\times$ faster (322\,s vs.\ 3108\,s) and sustains over an order of magnitude higher training throughput (4.97 vs.\ 0.41 samples/s). This gap makes \ac{TM} appealing for rapid re-training when data distributions shift or when performing frequent on-device updates, with the hardware used in this study.

\begin{table}[t]
\centering
\caption{\ac{CTM} (incl. preprocessing) vs.\ \ac{CNN}.}
\label{tab:tm-cnn}
\small
\begin{tabular}{lcc}
\toprule
\textbf{Metric} & \textbf{\ac{TM}} & \textbf{\ac{CNN}} \\
\midrule
Training Time (s) & 321.69 & 3107.91 \\
Inference Throughput (samples/s) & 46.23 & 78.23 \\
Model Memory Usage (MB) & 45.06 & 623.69 \\
\bottomrule
\end{tabular}
\end{table}

\paragraph{Inference cost}
\ac{CNN} inference is faster on this workload (5.11\,s vs.\ 8.65\,s total), yielding higher inference throughput (78.23 vs.\ 46.23 samples/s). If peak online throughput is the primary constraint, \ac{CNN} currently holds an advantage.

\paragraph{Memory footprint}
\ac{CTM} uses $\sim$14$\times$ less model memory (45.06\,MB vs.\ 623.69\,MB), which directly benefits embedded targets and eases model distribution/updates.

\paragraph{Explainability}

\begin{figure}[!htbp]
\centering
\includegraphics[width=0.7\linewidth]{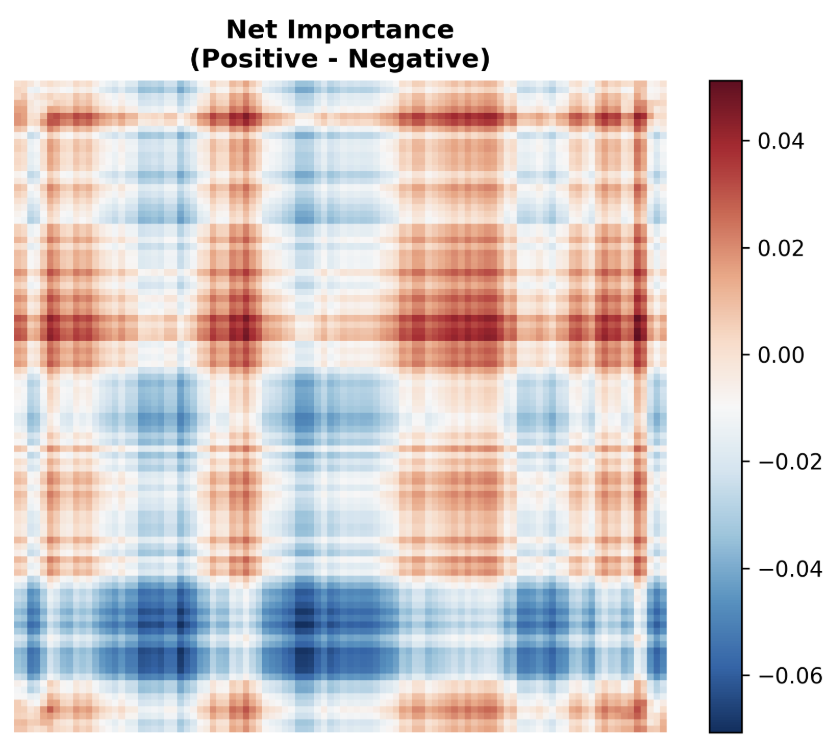}
\caption{Contributions of individual pixels, Red - Jamming, Blue - Pure, according to the trained \ac{CTM}, using literal mentions count}
\label{fig:CTM_posneg}
\end{figure}

\begin{figure}[!htbp]
\centering
\includegraphics[width=0.7\linewidth]{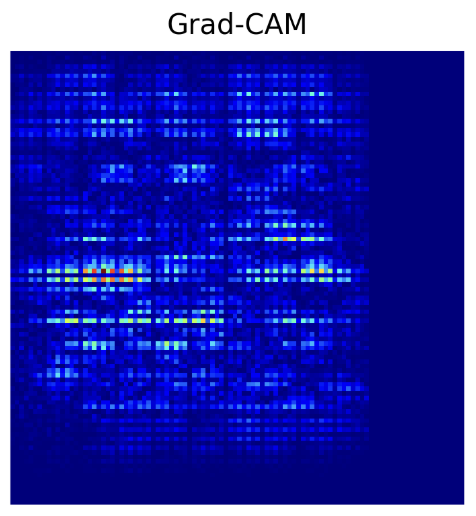}
\caption{Contributions of individual pixels, Red - Jamming, Blue - Pure, according to the trained \ac{CNN}, using GradCAM} 
\label{fig:CNN_GradCAM}
\end{figure}

In Fig. \ref{fig:CTM_posneg} and Fig. \ref{fig:CNN_GradCAM}, we show the difference between features used for the classification of each model. The \ac{CNN} seems to ignore lower frequency bands and focuses on specific parts of the spectrograph. Meanwhile, the \ac{CTM} uses the lower frequency bands, in almost full time slot of the spectrograph to consider the sample as Pure. Same as in our preprocessed dataset, each pixel in the 100×100 spectrogram represents approximately 78.7 kHz of frequency bandwidth from the original 15.625 MHz sampled signal. This can further be traced to the corresponding Resource Element.

\paragraph{Takeaways}
\ac{CNN} offers higher accuracy and faster inference, while \ac{TM} is more training-efficient and compact. CNNs are a better fit for latency-critical tasks with high performance hardware; TMs fit constrained or frequently retrained setups. Future work will explore lightweight CNNs and TM–CNN hybrids to combine strengths. Based on prior \ac{FPGA} studies and logic-sharing methods (e.g. MATADOR), we estimate our \ac{CTM} resource use on Z7 without new measurements. Assuming ~30–60 \acp{LUT} per specialized clause with shared logic and RAM-stored \acp{TA}, Table \ref{tab:fpga_profiles} shows estimated deployment profiles.

\begin{table}[t]
\centering
\caption{Projected \ac{CTM} inference footprints on \ac{Z7}. \ac{LUT}s use a 30--60/clause envelope, 100~MHz, stride~=~1, and 80~\% efficiency. No new measurements; \textbf{literature-based projection only.}}
\label{tab:fpga_profiles}
\small
\begin{tabular}{lcccc}
\toprule
\textbf{Profile} & \textbf{Clauses} & \textbf{\ac{LUT}s} & \textbf{Samples/s}\\
\midrule
Power & 256 & $\sim$7.7k--15.4k & $\sim$1.0--2.5k \\
Latency & 512 & $\sim$15k--31k & $\sim$2.0--4.6k \\
Accuracy & 800 & $\sim$24k--48k & $\sim$1.1--2.5k \\
\bottomrule
\end{tabular} \vspace{-3mm}
\end{table}

\section{Conclusion} \label{Sec:Conclusion}

This work introduced an explainable, hardware-efficient \ac{CTM} as an edge deployable alternative to a \ac{CNN} for jamming detection from 5G \ac{SSB} features. If \emph{maximizing detection accuracy and peak throughput} is the primary goal on general-purpose compute, the \ac{CNN} is preferable (accuracy $=96.83\pm1.19$). If meeting tight edge constraints such as small memory, low power, predictable latency, and interpretability, is a priority, the \ac{CTM} is advantageous: it maintained competitive performance (accuracy $=91.53\pm1.01$), trained $\sim\!9.5\times$ faster, and required $\sim\!14\times$ less model memory (45~MB vs.\ 624~MB). Literature-grounded \ac{FPGA} projections further indicate that clause-specialized \ac{CTM} designs fit within modest \ac{LUT} envelopes on \ac{Z7} and can deliver kSamples/s-class effective throughput at 100~MHz, enabling fully on-chip, low-power deployment. In short: \emph{choose \ac{CNN} for absolute accuracy; choose \ac{CTM} for explainable, resource-efficient edge inference}.

We are currently exploring hybrid CTM/CNN models and hardware-in-the-loop evaluations to optimize detection latency and resilience in dynamic RF environments, aiming to improve jamming defense in 5G and support progress toward B5G.

\section*{Acknowledgment}
This work is supported in part by the Norwegian Directorate for Higher Education and Skills under Project Number UTF-2021/10133, and in part by the Natural Sciences and Engineering Research Council of Canada (NSERC) DISCOVERY and CREATE TRAVERSAL Programs.
\bibliographystyle{IEEEtran}
\bibliography{refs}


\end{document}